\documentclass[twocolumn,letter]{jpsj2} 
\usepackage{bm}

\title{Magnetic Excitation in Artificially Designed Oxygen Molecule Magnet}

\author{Takatsugu \textsc{Masuda}\thanks{E-mail: tmasuda@yokohama-cu.ac.jp}, Satoshi \textsc{Takamizawa}, Kazuma \textsc{Hirota}$^{1}$, 
Masaaki \textsc{Ohba}$^{2}$, and Susumu \textsc{Kitagawa}$^{2}$}

\inst{International Graduate Schools of Arts and Sciences 
Yokohama City University, Yokohama, 236-0027 \\
$^{1}$Institute of Solid State Physics, the University of Tokyo, Kashiwa, 277-8581 \\
$^{2}$Department of Synthetic Chemistry and Biological Chemistry, Kyoto University, Kyoto, 615-8510}

\abst{We performed inelastic neutron scattering experiment to 
study magnetic excitation of O$_2$ molecules adsorbed in 
microporous compound. 
The dispersionless excitation with characteristic 
intensity modulation is observed at $\hbar \omega = 7.8$ meV 
at low temperature. 
The neutron cross section is explained by 
spin dimer model with intradimer distance of 3.1 \AA . 
Anomalous behaviour in the temperature 
dependence is discussed in the context of enhanced magnetoelasticity 
in the soft framework of O$_2$ molecule. }

\kword{Oxygen molecule, spin dimer, neutron scattering}

\begin{document}
\maketitle

Oxygen molecule, the second most abundant constituent in air, is essential 
to human beings. One of the interesting physical properties is that the 
diatomic molecule has magnetism with effective spin of $S$ = 1. In solid 
state at low temperature the oxygen behaves as insulating low-dimensional 
antiferromagnets\cite{Uyeda} thorough exchange interaction induced by the significant 
overlap of the molecular orbital. Under pressure the change of electronic 
structure brings exotic phenomena such as metallization\cite{Desgreniers}, 
superconductivity\cite{Shimizu} 
and exotic magnetic orders\cite{Freiman,Goncharenko}. 
Now the understanding of magnetic oxygen as 
an elementary substance has been progressed. The next challenge is to 
design a magnet by treating oxygen molecule as a magnetic entity. 
Enterprising trial for the O$_2$ based magnet was performed in adsorbing 
O$_2$ monolayer on graphite substrate\cite{McTague,Murakami}. 
Neutron diffraction and magnetic 
susceptibility study revealed the structural and magnetic variation of 
the artificially designed triangular lattice of O$_2$ molecule with the 
change of the oxygen density and temperature. However the amount of 
the oxygen in the monolayer is too small to be detected for a dynamical 
probe such as neutron inelastic scattering technique. For further 
magnetic study availability of bulk size magnet has been required. 

\begin{figure}
\includegraphics{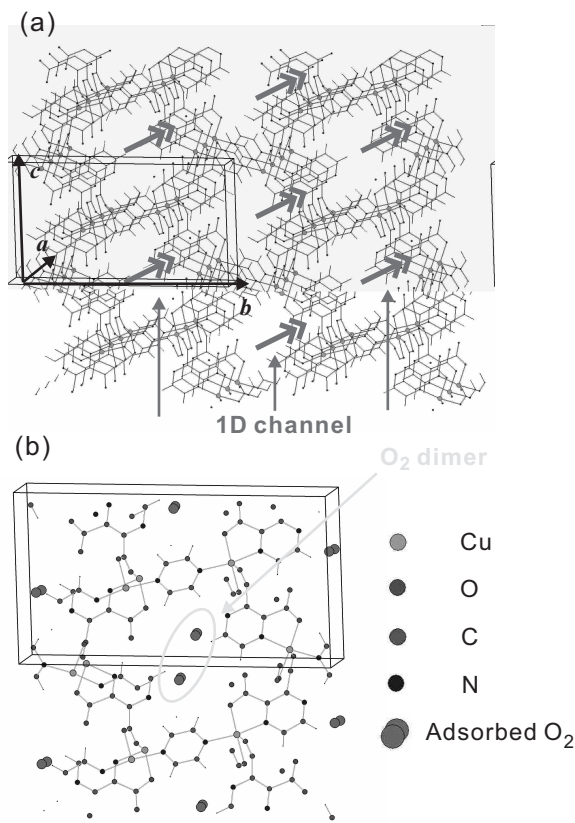}
\caption{\label{fig1} (Color online) 
Crystal structure of the microporous copper 
coordination polymer (CPL-1). 
(a) The framework of non-adsorbed CPL-1. 1D channel for molecules adsorption 
runs in crystallographic a direction. (b) Structure of CPL-1 with O$_2$ 
adsorption. }
\end{figure}

One solution is to utilize microporous compounds in a recently 
developing chemistry area, coordination polymers 
science\cite{Kitagawa04Angew,Mori97,Takamizawa04Angew}. Particularly 
interesting is a copper coordination polymer 
$\{$[Cu$_2$(pyrazine-2,3-dicarboxylate)$_2$(pyz)]2H$_2$O$\}_n$ abbreviated 
as CPL-1\cite{Kondo99Angew}. 
Molecule formula is 
C$_{16}$H$_{12}$Cu$_{2}$N$_{6}$O$_{10}$. 
The space group is monoclinic $P21/c$ and the lattice constants are 
$a$ = 4.693 \AA , $b$ = 19.849 \AA , $c$ = 11.096 \AA , and $\beta$ = 
96.90$^{\circ}$. 
The compound includes 1D channel for physical adsorption 
of molecules with the dimension of 4.0 \AA\ by 6.0 \AA\ in crystallographic 
$a$ direction (Fig. \ref{fig1}(a)). 
A strong confinement effect due to the restricted 
bore size enables the formation of ladder-like array of adsorbed 
oxygen\cite{Kitaura02Science} thorough van der Waals interaction. 
The intermolecular distance 
in the rung/dimer direction is 3.28(4) \AA . Unlike the surface adsorption 
on graphite O$_2$ molecules penetrate inside and the amount of the adsorption 
is one molecule per one copper atom. Hence the O$_2$ based magnet of bulk size 
is realized. The magnetic susceptibility and magnetization measurements 
suggest a non-magnetic ground state with a spin 
gap\cite{Kitaura02Science,Kobayashi05}. The magnitude of 
the gap was roughly estimated as 88 K $\sim$ 100 K (7.6 meV $\sim$ 8.6 meV). 
In this letter we report 
the first observation of the magnetic excitation in the prototype of 
artificial O$_2$ magnet realized in CPL-1 by using inelastic 
neutron scattering technique. 

Neutron scattering experiment was performed using PONTA spectrometer installed
 in 5G beamline in JRR-3M reactor in Japan Atomic Energy Agency. A series of
 scans were collected by changing incident neutron energy for fixed monitor
 counts in a low-efficiency detector between pyrolytic graphite (PG (002))
 monochromator and sample. 80' sollar collimator was used between
 monochromator and sample. Radial collimator was installed between
 sample and PG horizontally focused analyzer. The energy of the final
 beam is fixed at 14.7 meV. PG filter was placed after sample to eliminate
 higher harmonics.  Projected full-width at half-maximum energy resolution
 of this configuration at elastic position is 0.96 meV. 

\begin{figure}
\includegraphics*[width=8.7cm]{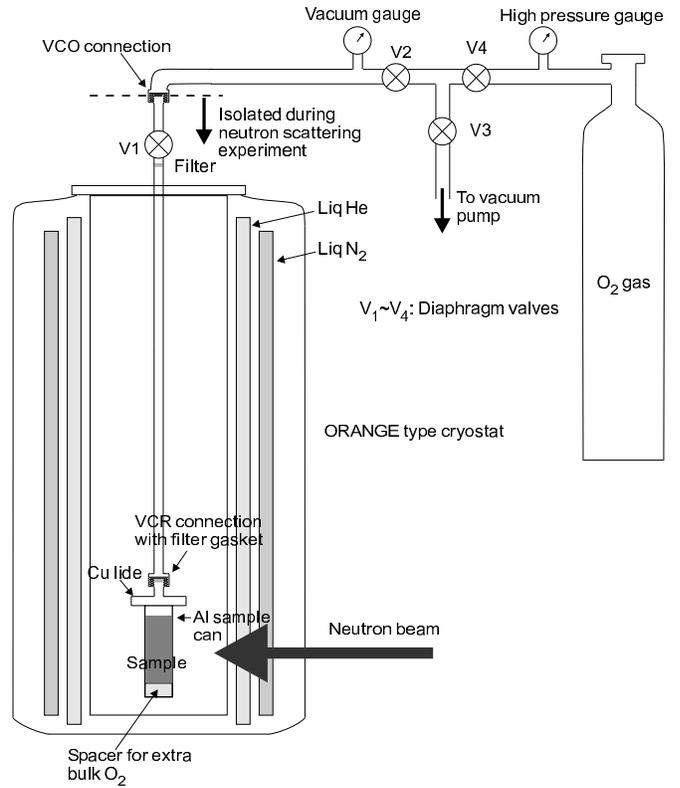}
\caption{\label{O2induction} Schematic view of O$_2$ induction apparatus. }
\end{figure}

Schematic view of O$_{2}$ induction apparatus is 
shown in Fig.~\ref{O2induction}. 
ORANGE type cryostat is used to
 achieve $T$ = 2.0 K. 12 g of powder sample was prepared by the previously
 reported method\cite{Kondo99Angew}. 
The sample is packed in Al cylindrical can 
 (20 mm in diameter and 67 mm in length) which can be attached to
 O$_2$ induction probe designed for the cryostat. 
VCR fitting with filter gasket (Swagelok Co. Ltd.) is used for 
the connection to stainless made O$_2$ induction tube. 
Spacer with the height of 1cm is put in the bottom of the Al can 
for the reservoir of extra bulk O$_2$. 
The adsorbed water 
molecules in the microporous channel are eliminated by heating 
the sample up to 400 K in reduced pressure. 

Since it is difficult to synthesize the deuterated CPL-1 
we use protonated sample. 
To measure background from the host CPL-1 including large 
incoherent scattering of hydrogen atoms 
we performed precedent scans in the non-adsorbed CPL-1 
for all scans. 
Afterwards the measurements for the O$_2$ adsorbed CPL-1 were performed. 
O$_2$ adsorption is processed by warming up the sample 
and by keeping the temperature at 95 K for an hour 
with O$_2$ pressure of 80 kPa. 

\begin{figure}
\includegraphics*[width=8.7cm]{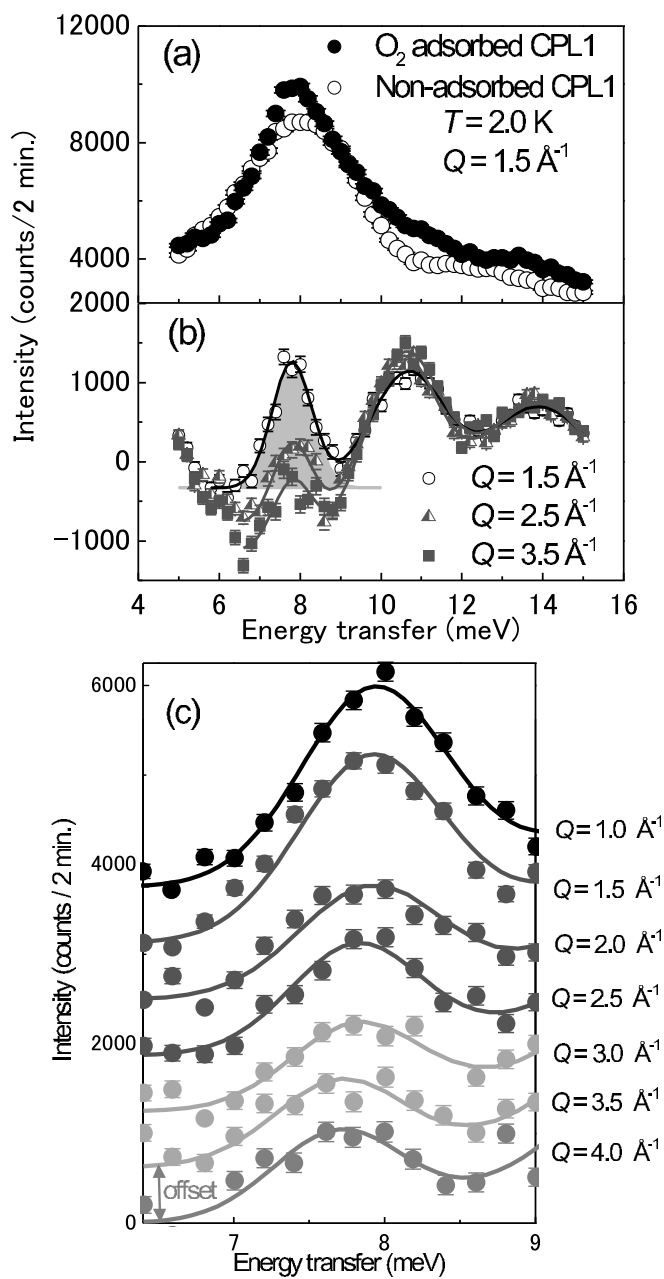}
\caption{\label{fig2} Excitations of O$_2$ based magnet from inelastic 
neutron scattering measurements at $T$ = 2.0 K. (a) Constant $Q$ scans 
at $Q$ = 1.5 \AA $^{-1}$ in O$_2$ adsorbed CPL-1 
(filled circles) and non-adsorbed CPL-1 
(open circles). (b) The subtracted intensity. 
Shaded area is the experimental resolution. 
(c) Net magnetic excitations in narrow energy range at 
various $Q$'s (offset for clarity).}
\end{figure}

Typical constant wave number ($Q$) scans of O$_2$ adsorbed and non-adsorbed 
CPL-1 are shown in Fig. \ref{fig2} (a). 
The contribution of the adsorbed oxygen is 
detected as the enhanced intensity at the energy transfer ($\hbar \omega$) 
of about 7.8, 10.6 and 13.7 meV. 
We assume that the intensity in O$_2$ adsorbed sample is, 
as zeroth approximation, simple sum of 
the excitation from the adsorbed O$_2$ and 
the background from the host compound. 
Then the subtracted excitation of the adsorbed oxygen 
is shown in Fig. \ref{fig2} (b). 
Solid curves are the fit to the 
data by triple Gaussian functions. 
Among the observed three peaks the peak at $\hbar \omega =$ 7.8 meV 
shows typical behaviours of the 
excitation of a magnetic cluster: the peak energy does not change with $Q$ 
and the peak intensity decreases with the increase of $Q$. 
In addition the peak width is within resolution limit as is indicated by 
shaded area. 
\begin{figure}
\includegraphics*[width=8.7cm]{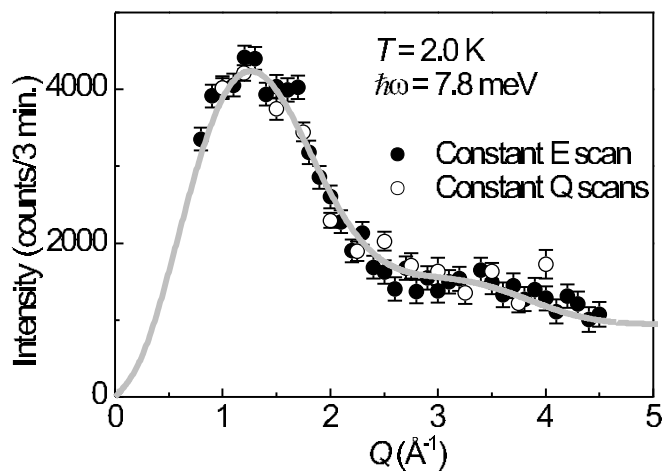}
\caption{\label{fig3} 
Wave number dependence of the neutron cross section of 
the adsorbed magnetic O$_2$. Symbols are the subtracted data and 
solid curve is the calculation of $S$ = 1 spin 
dimer model in eq. (1) in the text. }
\end{figure}
On the other hand the intensity of the peaks at $\hbar \omega = 10.6$ and 
13.7 meV is almost $Q$ independent. 
These peaks can be ruled out from the magnetic origin since a 
magnetic neutron cross section is reduced at high $Q$ due to magnetic form 
factor\cite{Lovesey}. 
For some unknown reasons the subtracted intensity has negative values 
at 6meV $\lesssim \hbar \omega \lesssim$ 10meV and we presumed negative 
background for the Gaussian fittings. 
Probably O$_2$ adsorption suppresses the incoherent 
scattering of hydrogen in host compound. 
The energy scans after the negative background subtraction are shown in 
Fig. \ref{fig2} (c). 
The characteristic behaviour of the magnetic cluster is observed more clearly. 
$Q$ dependence of nuetron cross section at $\hbar \omega$ = 7.8 meV 
is shown in Fig.~\ref{fig3}. 
Filled circles are the net magnetic excitation of adsorbed O$_2$ 
obtained by similar data analysis in Fig.~\ref{fig2}(c). 
Open circles are 
the scaled intensity of constant $Q$ scans obtained in Fig. \ref{fig2}(c). 

The most probable spin model for the magnetic cluster is Heisenberg $S$ = 1 
spin dimers from crystallographic consideration\cite{Kitaura02Science}. 
The effective spin 
Hamiltonian is described by $H = -2J {\bm S}_1 \cdot {\bm S}_2$. The calculated eigenenergies are $E(S) = -J\left( S \left( S+1 \right) -3/2 \right)$ where 
$S$ 
is the magnitude of the total spin operator ${\bm S} = {\bm S}_1 + {\bm S}_2$ 
($S$ = 0, 1, 2). Each energy 
level is $2S+1$ - fold degenerated and 
the energy levels satisfy Lande-interval 
rule. The magnetic excitation at low temperature is dominated by the 
transition between non-magnetic singlet ground state and $S$ = 1 
triplet state. 
The neutron cross section for the magnetic scattering is obtained\cite{Furrer} 
\begin{equation}
\frac{d^2 \rho}{d\Omega d\omega} \propto p_1 (T)F^2(Q)\left(1-\frac{\sin (QR)}{QR}\right) \delta (\hbar \omega + 2J).
\label{eq1}
\end{equation}
Here $p_1 (T)$ is a temperature factor that is proportional to 
the thermal population 
of singlet ground state: $p_1(T)=1/\left(1+3\exp (2J/k_BT)+5\exp (6J/k_BT)\right)$. $F(Q)$ is magnetic form factor of oxygen molecule, $Q$ is a 
scattering wave number and $R$ is intradimer 
distance between oxygen molecules. 
For the calculation of $F(Q)$ the density of atomic $P\pm$ 
orbitals is assumed for 
the spin density of an O$_2$ molecule.\cite{Kleiner} 
Then it is obtained that 
$F({\bm Q})=\int_0^{\infty }dr\int_0^{\pi}d\theta r^4 \sin ^3 
\theta e^{-2br^2}[{\rm cosh}(2brR_0 \cos \theta)-1]
e^{iQr\cos (\theta - \beta)}.$
Here $R_0$ = 1.21 \AA\ and $b$ = 4.1 \AA $^{-2}$~\cite{Stephens}. 
$R_0$ is the internuclear distance, $b$ is a fitting parameter, and 
$\beta$ is the angle between molecular axis and scattering vector ${\bm Q}$. 
$F(Q)$ is numerically obtained by powder average $F(Q)=\int F({\bm Q}) d\Omega$. 

\begin{figure}
\includegraphics{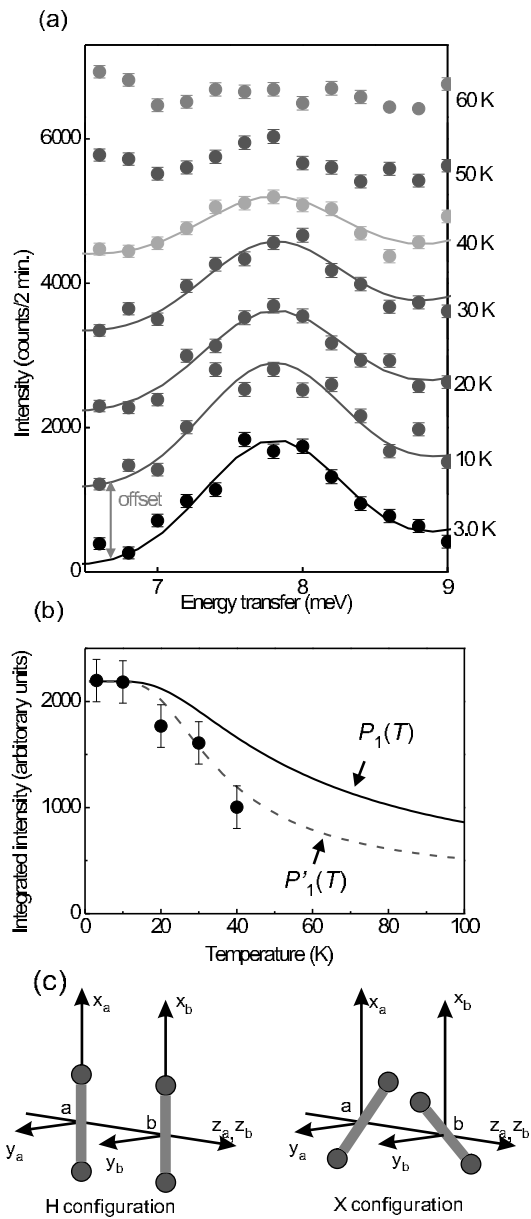}
\caption{\label{fig4} Temperature dependence of the magnetic excitation. 
(a) Constant wave-vector scan at $Q$ = 1.5 \AA $^{-1}$ at various temperatures 
(offset for clarity). Solid curves are guide for the eyes. (b) Temperature 
dependence of the integrated intensity. Solid and dashed curves are 
temperature factors described in the text. (c) O$_2$ dimer in H configuration 
(left) and X configuration (right). }
\end{figure}

The $Q$ independent magnetic excitation 
at 7.8 meV in Fig. \ref{fig2}(c) is 
consistent with eq. (\ref{eq1}) with $2J = -7.8$ meV. 
Characteristic feature of the dimers 
neutron cross section is identified in constant energy scan at 
$\hbar \omega = 7.8$ meV in 
Fig. \ref{fig3}. 
Enhanced peak structure at $Q \sim 1.3$ \AA $^{-1}$ 
and drastic suppression of the 
intensity with the increase of $Q$ are perfectly reproduced by the calculation 
based on $S$ = 1 spin dimer model with $R$ = 3.1 \AA . The value of $R$ 
is reasonably 
consistent with intermolecular distance of $R$ = 3.28 \AA\ 
obtained by synchrotron 
x-ray diffraction\cite{Kitaura02Science}. 
These facts mean that magnetic excitation at $T$ = 2.0 K is 
explained by the adsorbed O$_2$ dimers. 
Furthermore, the energy scale is consistent with the spin 
gap estimated by bulk measurements 
7.6 $\sim$ 8.6 meV.\cite{Kitaura02Science,Kobayashi05}

Constant Q scans at various temperatures is shown in Fig. \ref{fig4}(a). 
The magnetic 
peak at $\hbar \omega = 7.8$ meV is suppressed with the temperature and 
almost disappears at $T \gtrsim 50$ K. 
As is shown in Fig. \ref{fig4}(b) 
the peak intensity goes down more rapidly than the 
temperature factor $p_1(T)$ shown by the solid curve. The data at 
$T \lesssim 40$ K is rather 
consistent with two level system with singlet ground and octet excited states 
described by an empirical 
formula $p_1'(T)=1/\left( 1+8\exp (2J/k_BT)\right)$.

Since the framework of O$_2$ molecules formed by van der Waals interaction is 
flexible, a magnetoelastic effect can be enhanced enough to affect the energy 
spectrum of the effective spin model. The calculation of the intermolecular 
potential of O$_2$ dimer showed that the energy level, particularly in high 
energy range, depends both on spin state and geometrical 
configuration\cite{Bussery}. 
The ground state is $S$ = 0 singlet with H configuration in 
Fig. \ref{fig4} (c) and the 
first excited state is $S$ = 1 triplet with the same H configuration. However, 
the second excited state is $S$ = 2 quintet with 
X configuration and the energy 
interval from the first excited state is roughly $-2J$. 
This value is two times 
smaller than that by Lande-interval rule of the spin dimer model. These 
calculations indicate that the effective spin Hamiltonian works only for the 
low energy excitation and the inclusion of van der Waals potential is 
essential to the higher energy excitations. For the understanding in 
quantitative level the consideration of specified potential including 
the interaction between O$_2$ and CPL-1 is required in the calculation. 

In summary we report the magnetic excitation of 
artificial O$_2$ magnet in microporous compound CPL-1 by using 
inelastic neutron scattering technique. 
While the neutron cross section 
at low temperature is completely explained by existent spin model, 
the temperature dependence shows anomalous behaviour due to enhanced 
magnetoelastic effect of the adsorbed O$_2$ molecules. 
Our study demonstrates that the conventional dynamical probe for physics 
study, inelastic neutron scattering technique, is powerful tool 
for an O$_2$ based magnet in recent developing chemistry area. 

This work was partly supported by Yamada Science Foundation, 
Asahi glass foundation, 
the Strategic Research Project (No. 19042) of Yokohama City 
University and Grant-in-Aid for Scientific Research 
(No.s 19740215 and 19052004) of Ministry of Education, 
Culture, Sports, Science and Technology of Japan.



\end{document}